\documentclass[preprint]{aastex701}
\usepackage{amsmath, graphicx, xcolor}

\newcommand{\com}[1]{} 

\newcommand{\figref}[1]{Fig.~\ref{#1}}

\newcommand{\secref}[1]{Sec.~\ref{#1}}
\newcommand{\sectionref}[1]{Section~\ref{#1}}
\newcommand{\eqnref}[1]{Eq.~\eqref{#1}}

\newcommand{\CAi}{C_{\text{Ai}}}
\newcommand{\CAe}{C_{\text{Ae}}}
\newcommand{\Cp}{C_\text{p}}
\newcommand{\Cs}{C_\text{s}}

\newcommand{\diff}[2]{\frac{\text{d}#1}{\text{d}#2}}

\newcommand{\pdiff}[2]{\frac{\partial #1}{\partial #2}}
\newcommand{\myexp}[1]{\text{e}^{#1}}
\newcommand{\fromto}[2]{#1\,\text{-}\,#2}

\newcommand{\mydeleted}[1]{} 

\graphicspath{.}

\begin{document}

\title{FAST Observations of Wave-like Structures in the Radio Dynamic Spectrum of AD Leo}

\author[orcid=0009-0001-6464-6519]{Wenjie Zou}
\affiliation{School of Earth and Space Sciences, Peking University, Beijing, 100871, China}
\email{2501110622@stu.pku.edu.cn} 

\correspondingauthor{Hui Tian}

\author[orcid=0000-0002-1369-1758]{Hui Tian}
\affiliation{School of Earth and Space Sciences, Peking University, Beijing, 100871, China}
\affiliation{State Key Laboratory of Solar Activity and Space Weather, National Space Science Center, Chinese Academy of Sciences, Beijing, 100190, China}
\email[show]{huitian@pku.edu.cn}  

\author[orcid=0009-0004-0713-405X]{Jiale Zhang} 
\affiliation{School of Earth and Space Sciences, Peking University, Beijing, 100871, China}
\affiliation{ASTRON, Netherlands Institute for Radio Astronomy, Oude Hoogeveensedĳk 4, Dwingeloo, 7991 PD, the Netherlands}
\affiliation{Kapteyn Astronomical Institute, University of Groningen, P.O. Box 800, 9700 AV, Groningen, the Netherlands}
\email{jialezhang@pku.edu.cn}  

\author[orcid=0000-0002-6641-8034]{Yuhang Gao}
\affiliation{School of Earth and Space Sciences, Peking University, Beijing, 100871, China}
\affiliation{Centre for mathematical Plasma Astrophysics, Department of Mathematics, KU Leuven, Celestijnenlaan 200B bus 2400, B-3001 Leuven, Belgium}
\email{yh_gao@pku.edu.cn} 

\begin{abstract}

M-dwarf flare stars like AD Leo are laboratories for studying intense magnetic activities. The coherent radio bursts they produce are powerful probes of stellar coronal plasma and magnetic fields. In this study, we present high-resolution observations of AD Leo from the Five-hundred-meter Aperture Spherical radio Telescope (FAST) that reveal wave-like structures in its radio dynamic spectrum. The observations show trains of short-duration, narrowband sub-bursts where the central frequency, frequency drift rate, and flux density are all simultaneously modulated with a period of 1.53 s. Notably, modulation of the central frequency is approximately in-phase with that of the drift rate but roughly in anti-phase with that of the flux density. Furthermore, the amplitude of the frequency modulation grows with an e-fold timescale of 2.4 s. We interpret the observed sinusoidal frequency modulations as a possible signature of a magnetohydrodynamic (MHD) wave in the stellar corona. Our work provides a window into stellar coronal seismology and offers an opportunity to infer the local plasma environment via the MHD wave model. 

\end{abstract}

\keywords{\uat{Radio bursts}{1339} --- \uat{M dwarf stars}{982} --- \uat{Stellar coronae}{305} --- \uat{Stellar magnetic fields}{1610} --- \uat{Stellar coronal loops}{309} --- \uat{Solar coronal waves}{1995}}

\section{Introduction}

M-dwarf stars are presumably the most common type of stars in the galaxy \citep{Henry_2024}, and many, including the well-known M3 dwarf flare star AD Leo \citep{Kesseli_2019}, exhibit intense magnetic activity. This activity often generates powerful, coherent radio bursts that serve as unique probes of the physical conditions within stellar coronae, such as plasma properties and magnetic field structures \citep{Dulk_1985, Bastian_1990b}. Understanding the mechanisms that drive and modulate these emissions is therefore crucial for developing a complete picture of stellar atmospheric physics.

The radio bursts from AD Leo frequently display a wealth of fine structures. Observations often reveal numerous short-duration, narrowband sub-bursts that drift rapidly in frequency \citep{Gudel_1989,Bastian_1990a,AbadaSimon_1997,Osten_2006,Osten_2008,Zhang_2023,Zhang_2025,Zhang_2026}. Both plasma emission and Electron Cyclotron Maser (ECM) emission have long been considered as possible mechanisms for such coherent emission \citep[e.g.,][]{AbadaSimon_1997, Osten_2006}. A recent high-resolution observation now provides stronger evidence for ECM, favoring it as the mechanism that can more likely produce the observed narrowband, millisecond-scale sub-bursts \citep{Zhang_2023}. While the emission mechanism has received much research attention, the processes that produce these fine structures are far from being understood. If interpreted correctly, they could serve as powerful diagnostic tools. 

Quasi-periodic pulsations (QPPs) are commonly observed phenomena in both solar and stellar flares \citep{McLaughlin_2018, Zimovets_2021}. These pulsations are often interpreted as signatures of magnetohydrodynamic (MHD) waves, which can modulate energy release, electron dynamics, and plasma properties \citep{Nakariakov_2009, Nakariakov_2020}. This interpretation allows the use of QPP periods for coronal seismology, inferring plasma parameters like density and magnetic field strength \citep{Tian_2016, VanDoorsselaere_2016, Nakariakov_2024}. While MHD waves in the solar corona can be studied through imaging and spectroscopy, we cannot resolve them on distant stars. For stellar flares, the tentative evidence for such waves comes from QPPs in lightcurves from radio to X-ray bands \citep{Mitra-Kraev_2005, Srivastava_2013, Doyle_2018, Kowalski_2024}. The information from these lightcurves is limited, often making it difficult to uniquely identify the underlying mechanism. Consequently, although MHD oscillations are a suggested cause of those stellar QPPs, alternative explanations such as quasi-periodic magnetic reconnection typically cannot be ruled out \citep{Nakariakov_2009, Zimovets_2021}, highlighting the need for more direct diagnostics of the underlying plasma dynamics.

We report recent AD Leo observations from the Five-hundred-meter Aperture Spherical radio Telescope \citep[FAST;][]{Nan_2011,Jiang_2020} that reveal a wave-like envelope of the elementary sub-bursts. While previous studies have noted intra-train quasi-periodicity of sub-bursts \citep[e.g.,][]{Gudel_1989,Bastian_1990a,Zhang_2023,Zhang_2025}, our high-resolution observations are the first to unambiguously demonstrate a global, quasi-periodic modulation of entire sub-burst trains. We identify a distinct, wave-like train of sub-bursts where the central frequency, frequency drift rate, and flux density are all simultaneously modulated in a quasi-periodic manner, with a growing frequency modulation amplitude. We propose that the observed global modulations on the sub-burst trains are possibly caused by MHD waves propagating through the coronal magnetic flux tubes that host the radiation source. By analyzing the wave-like trains, we may constrain the properties of the MHD wave and the local coronal environment. 

In \sectionref{sec:obs}, we describe the FAST observations and the characteristics of a modulated sub-burst train. In \sectionref{sec:model}, we develop an interpretive framework based on the MHD wave model. Finally, we summarize our conclusions in \sectionref{sec:sum}.

\section{Observation}
\label{sec:obs}

AD Leo (GJ 388) is a nearby (4.97 pc) M3 dwarf with a mass of 0.41$M_\odot$, a radius of 0.435$R_\odot$ and a rotation period of 2.23 days \citep{Hunt-Walker_2012, Mann_2015, Kesseli_2019}. It is a magnetically active star, renowned for its frequent and powerful flares observed from X-ray to radio bands \citep[e.g.][]{Hawley_1995,Favata_2000,Gudel_2003,Chacon_2006,Hunt-Walker_2012,Stelzer_2022,Zhang_2024,Ram_2025}. 

We conducted observations of AD Leo with the FAST during four epochs: from 2023-11-29 20:05 to 2023-11-30 00:05, from 2023-11-30 20:50 to 2023-12-01 00:50, from 2023-12-01 19:50 to 2023-12-01 23:50, and from 2023-12-05 20:50 to 2023-12-06 00:50 (UT) \footnote{Publicly released FAST data (project code: PT2023\_0019). See \url{https://fast.bao.ac.cn/}.}. We employed the 19-beam receiver working at \fromto{1.0}{1.5} GHz in the L-band, achieving a temporal resolution of 0.4 ms (after $2\times$ time rebinning from the original $\sim$0.2 ms sampling) and a spectral resolution of $\sim$0.5 MHz per channel across 1024 uniformly spaced frequency channels. Data reduction followed the methodology established by \citet{Zhang_2023}, including Stokes parameter conversion, polarization/flux calibration via noise diode injection, and removal of slow-varying background trends. The final calibrated dynamic spectra represent flux density in mJy units in the four Stokes parameters $(I, Q, U, V)$.

\begin{figure}[]
    \centering
    \includegraphics[width=\textwidth]{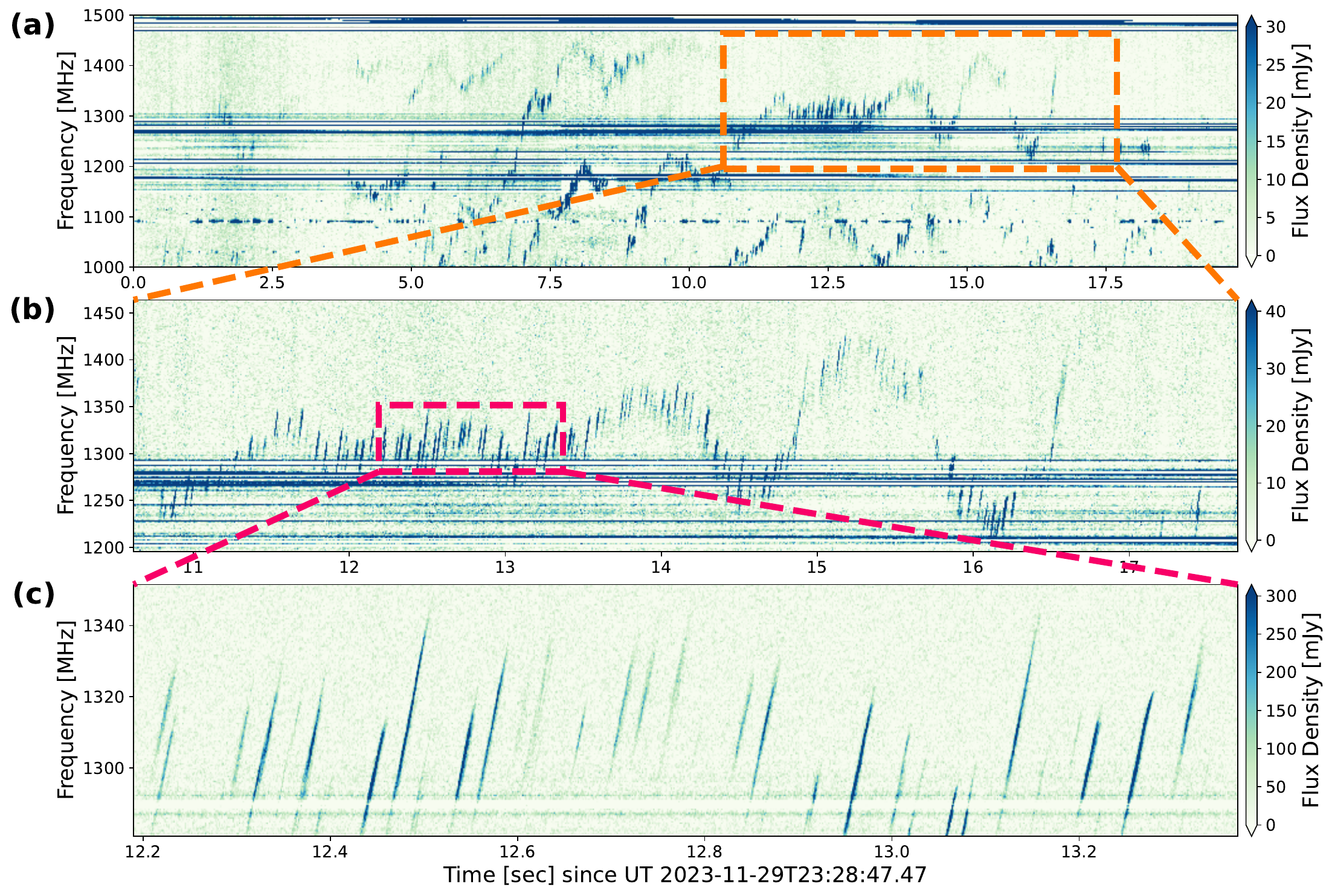}
    \caption{Dynamic spectrum (Stokes $I$ component) of the radio burst from AD Leo. (a) The radio burst exhibiting wave-like sub-burst trains. (b) Zoomed-in view of the sub-burst train marked in (a), showing distinct periodicity and amplitude growth. (c) Further zoom around a minor frequency modulation peak in (b), revealing the burst's linear sub-burst fine structures. Note the intensity decrease specifically at this wave peak phase. Horizontal streaks persisting throughout the time window represent radio frequency interference.}
    \label{fig:overview}
\end{figure}

The 16 hours of FAST observations, distributed across the four epochs, captured multiple radio burst events, revealing abundant fine structures in the dynamic spectra. One event on November 29 (UT) exhibited wave-like trains of linear sub-bursts, where the central radiation frequency underwent modulation, as shown in \figref{fig:overview}. This event consisted of several wave-like sub-burst trains with irregular morphologies, along with one wave-like train displaying distinct periodicity and amplitude growth. A detailed description of this event follows.

\begin{figure}[]
    \centering
    \includegraphics[width=\textwidth]{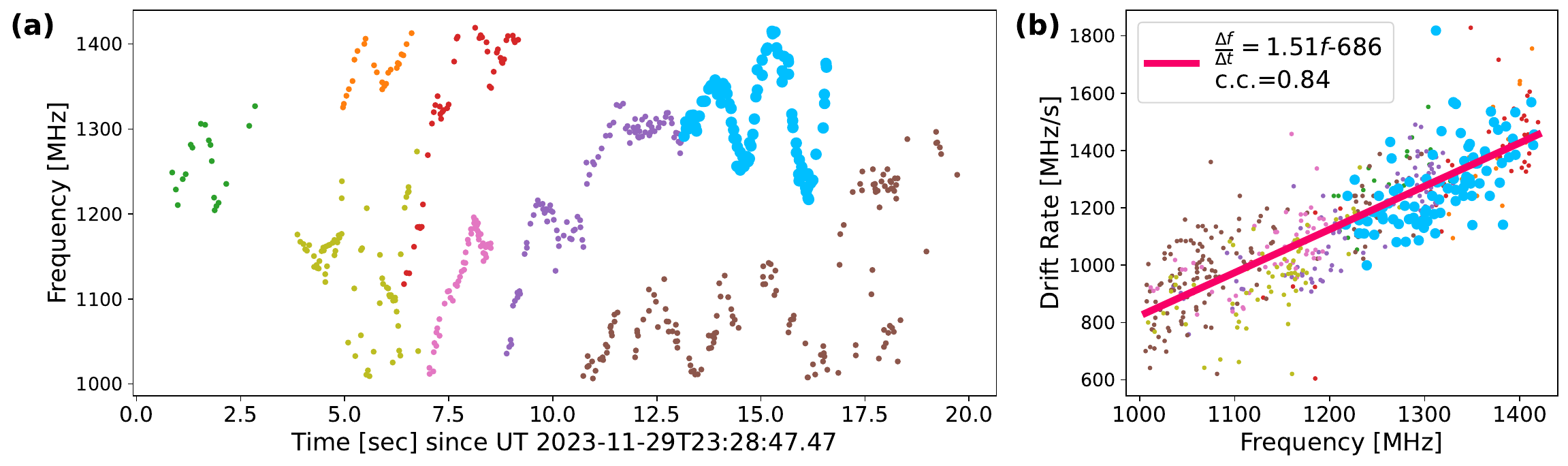}
    \caption{Results of the manual sub-burst start/end point annotation. 540 sub-bursts were annotated within the dynamic spectrum shown in \figref{fig:overview}(a). (a) Distribution of the annotated sub-bursts' central points. Larger light-blue points form the distinct wave-like train, and were used for curve fitting in \figref{fig:mod}(a). (b) Sub-burst drift rate versus central frequency, color-coded to match (a). The red line indicates a linear fit result.}
    \label{fig:label}
\end{figure}

\subsection{Basic Properties of the Sub-bursts}
\label{sec:subburst_obs}

The $\sim$20\,s radio burst in the observed band consisted of numerous short-duration, narrowband, rapidly drifting linear sub-bursts. Sub-burst start and end points were annotated manually (\figref{fig:label}(a)), and a heuristic algorithm was used to extract sub-burst pixels. The key properties of the sub-bursts are summarized below. 

The typical flux density of sub-burst pixels was $\sim$165\,mJy, occasionally exceeding 1\,Jy, compared with an rms noise level of $\sim$90\,mJy at 0.4\,ms resolution. The emission was highly circularly polarized, with a median degree of $V/I \approx 0.87$ (left-handed), and no detectable linear polarization. Instantaneous bandwidths were $\sim$5\,MHz ($\delta f/f \sim 0.3\%$), while frequency differences between start and end points ranged from 10 to 50\,MHz. At a single frequency, sub-bursts typically lasted for $\sim$5\,ms. Their total durations are $\sim$30\,ms. Their frequency drift rates were positive, lying between 800 and 1400\,MHz\,s$^{-1}$, and showed a strong positive correlation with central frequency (\figref{fig:label}(b)). Sub-bursts were distributed across nearly the entire observed band, often grouped into trains with a quasi-periodic intra-train spacing of $\sim$40\,ms. 

Similar sub-bursts have been reported in earlier AD Leo observations \citep{Gudel_1989,Bastian_1990a,AbadaSimon_1997,Osten_2006,Osten_2008,Zhang_2023,Zhang_2025}. Consistent with \citet{Osten_2008} and \citet{Zhang_2023}, we regard the features of individual sub-bursts as intrinsic to the source in AD Leo's corona, rather than artifacts from propagation effects, radio frequency interference, or other sources within the main beam.

\begin{figure}[]
    \centering
    \includegraphics[width=\textwidth]{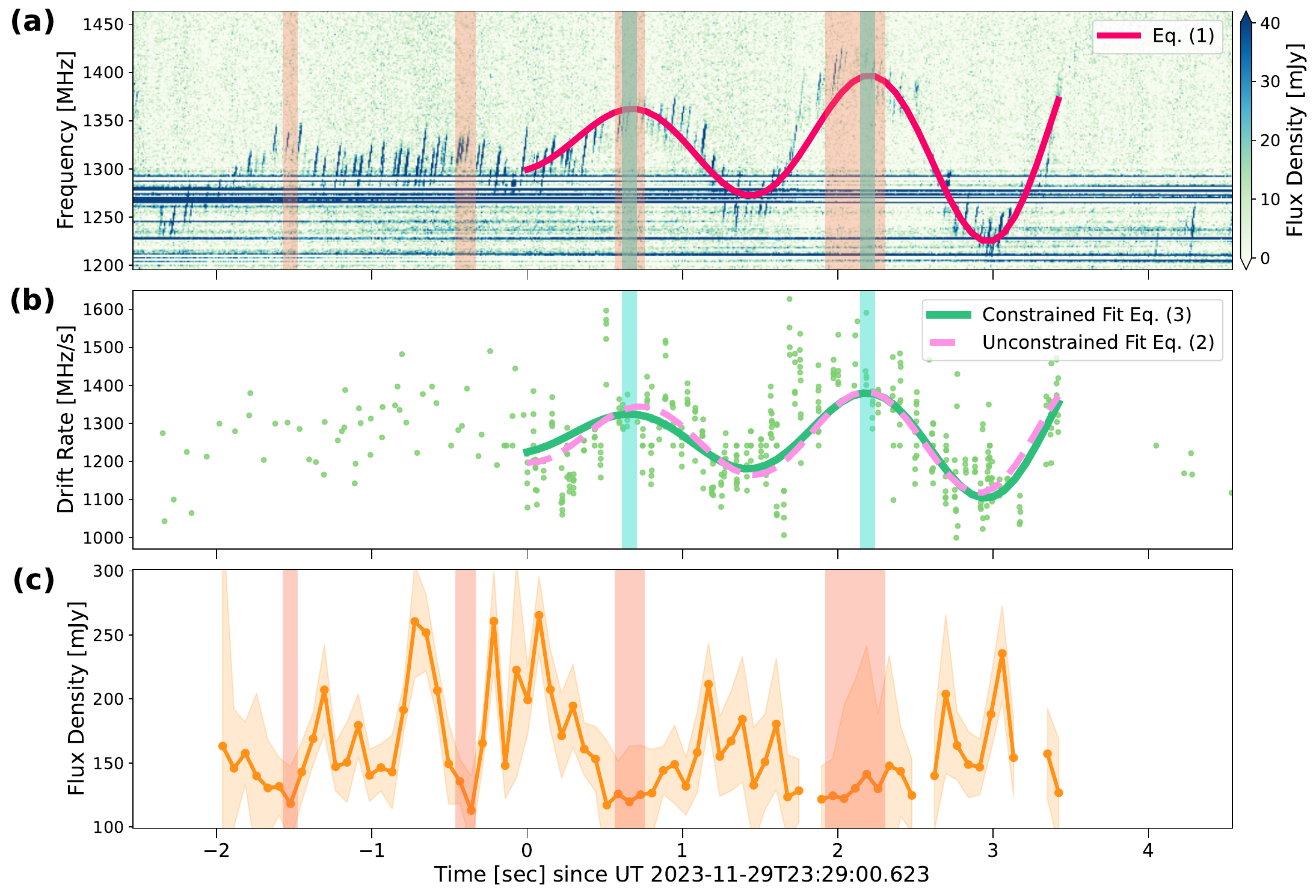}
    \caption{Modulation characteristics of the major sub-burst train. (a) The sub-burst train analyzed in \figref{fig:overview}(b). The red curve shows the fitted central frequency evolution \eqnref{eqn:freq_obs} for the distinctively periodic latter half. (b) Green points: drift rates derived from manually annotated start/end points, with the second half consisting of five rounds of repeated annotations. The pink dashed curve shows the unconstrained fitting result of \eqnref{eqn:drift_obs_free}, while the green curve shows the drift rate evolution \eqnref{eqn:drift_obs} from the constrained fit. Cyan vertical bars mark fitted peaks, corresponding approximately to frequency modulation peaks in (a). (c) Orange line: binned average flux density from algorithm-extracted pixels. The shaded region around the line represents the $\pm 2 \sigma$ uncertainty. Orange vertical bars mark approximate flux minima, roughly aligning with frequency modulation peaks in (a). Note: Time axis zero differs from \figref{fig:overview}; six outliers were omitted from (b) for visual clarity, though the fits incorporated the complete dataset.}
    \label{fig:mod}
\end{figure}

\subsection{Modulations of the Sub-burst Trains}
\label{sec:mod_obs}
The radio burst contained multiple sub-burst trains. Their central radiation frequencies oscillated on timescales of $\sim$1 second, forming wave-like structures. Some of the trains showed irregular structures such as an overall upward frequency drift. They were also generally spaced in frequency by a similar amount (roughly 200 MHz). As shown in \figref{fig:mod}, one train featured a less pronounced frequency modulation in its first half, while its second half exhibited distinct periodic modulation with increasing amplitude. After approximately 3.5 seconds (spanning two oscillation periods), the latter section faded below detectability. The following measurements characterize this specific train.

First, sub-bursts within the train were manually identified by annotating start/end points. We annotated the train's second half for five times, and fitted its central frequency using the least-squares method with an exponentially growing sinusoidal function:
\begin{align}
f=1325+28.5\,\myexp{t/2.35}\cos{\left(\frac{2\pi}{1.53}t-0.85\pi\right)}\,\text{MHz}. \label{eqn:freq_obs}
\end{align}
Hereafter, time $t$ is in seconds. The fit gives a relative modulation amplitude $\leq10\%$.

Given that drift rates remained approximately constant during individual sub-bursts, they could be calculated from the start/end annotations. The green points in \figref{fig:mod}(b) demonstrate that the drift rate also underwent a quasi-periodic modulation. Fitting the drift rates with an exponentially growing sinusoidal functional form with all five parameters free yielded:
\begin{align}
\diff{f}{t}=1263+67.0\,\myexp{t/3.73}\cos{\left(\frac{2\pi}{1.45}t-0.99\pi\right)}\,\text{MHz/s}. \label{eqn:drift_obs_free}
\end{align}
This fitting result is depicted by the pink dashed curve in \figref{fig:mod}(b). We noticed the period and the e-fold growth time deviate from those of the frequency modulation. We hypothesized that the discrepancies could arise from human-induced bias in labeling and modeling. Based on a physical reasoning that both modulations should share the same underlying oscillation parameters, we fixed the period and e-fold growth time to the values from the frequency modulation. This yielded a constrained fit:
\begin{align}
\diff{f}{t}=1264+46.0\,\myexp{t/2.35}\cos{\left(\frac{2\pi}{1.53}t-0.83\pi\right)}\,\text{MHz/s}. \label{eqn:drift_obs}
\end{align}
This result is illustrated by the green curve in \figref{fig:mod}(b), which almost overlaps with the unconstrained fitting result. Despite some differences in scalar values of the fitting results with different fitting strategies, we can still conclude that the frequency modulation is approximately in phase with the drift rate modulation. 

In addition to frequency and drift rate modulation, the flux density of sub-bursts within the train also showed quasi-periodic modulation. Using the heuristic sub-burst pixel extraction algorithm, the sliding-window average flux density of the sub-bursts is obtained and shown by the orange curve in \figref{fig:mod}(c). Flux minima (orange vertical bars) generally coincide temporally with frequency maxima, suggesting an approximate anti-phase relationship between flux density and central frequency modulation. The modulation depth $(I_\text{max}-I_\text{min})/(I_\text{max}+I_\text{min})$ could reach $\sim$35\%; however, because the pixel extraction algorithm systematically overestimates the average intensity of weak sub-bursts, the actual depth may be greater. Near a minor wave peak at $\sim$$-0.5$ s in \figref{fig:mod}(a), sub-burst flux density decreased significantly. Zooming in the dynamic spectrum around this peak (\figref{fig:overview}(c)) revealed no clear intensity decrease with increasing frequency during each individual sub-burst drift, and reduced flux density occurred only in the vicinity of this specific wave peak. This indicates that the radiation intensity is tied to the phase of the wave-like structure rather than merely exhibiting an anti-correlation with frequency. Although modulation amplitude of the central frequency increased over time, flux density modulation showed no matching amplitude trend. Similar calculation revealed no significant evidence of circular polarization degree modulation. 

\citet{AbadaSimon_1997} reported second-scale frequency variations of sub-bursts in radio dynamic spectra of AD Leo. We also noticed that the central frequency of the sub-bursts varied in Figure 1(c) of \citet{Zhang_2023}. Analyzing an AD Leo burst light curve of $\fromto{4.5}{5.1}\,$GHz, \citet{Zaitsev_2004} attributed their observations to a $\sim$5 second periodic intensity modulation of pulsed radio emission. Compared with these studies, the present FAST observation is the first to unambiguously reveal a clear, sinusoidal envelope modulating the sub-burst train's central frequency. With high time-frequency resolution, we further demonstrate the simultaneous modulation of the drift rate and the flux density.

\section{Discussion} 
\label{sec:model}

\subsection{The Emission Mechanism of Individual Sub-bursts}
\label{sec:ecm_theory}
The narrowband, highly circularly polarized and rapidly drifting sub-bursts observed in the dynamic spectrum of AD Leo are primarily interpreted as ECM emission \citep{Zhang_2023}. ECM is a well-established coherent radiation mechanism for stellar and planetary radio emissions, driven by a non-thermal electron velocity distribution, such as a loss-cone, in a strong magnetic field \citep{Dulk_1985,Treumann_2006,melrose2017coherent,Callingham_2024}. The emission typically occurs at the fundamental or second harmonic of the electron cyclotron frequency, which is directly proportional to the local magnetic field strength: $f_{\text{ECM}} = s f_{\text{ce}} = 2.80\,sB\,\text{MHz}$, where $B$ is the magnetic field in Gauss and $s=1,2$ is the harmonic number. The detected frequency ($\fromto{1.0}{1.5}\,$GHz) corresponds to a magnetic field strength of $B=\fromto{357}{536}\,\text{G}$ at the source location for fundamental emission ($s=1$), and $B=\fromto{179}{268}\,\text{G}$ for the second harmonic ($s=2$). Furthermore, the ECM mechanism requires a low plasma density where the plasma frequency is less than or similar to the electron gyro-frequency \citep{Treumann_2006, Lu_2025}. This implies a plasma number density $n<10^{10.4}\,\text{cm}^{-3}$ (estimated by using an upper bound plasma frequency of 1.5 GHz) and a low plasma beta condition ($\beta \ll 1$). We note that the potential issue of gyroresonance absorption at harmonic layers for ECM emission escape has been discussed for AD Leo \citep[e.g.,][]{Stepanov_2001,Osten_2006,Osten_2008}. The ECM interpretation remains plausible, as propagation effects or a low density at harmonic layers may mitigate significant absorption, and the original high flux density may remain detectable even with some attenuation \citep{Bastian_1990a, Ergun_2000,Vedantham_2021, Yu_2024}. As detailed by \citet{Zhang_2023}, the rapid frequency drift is naturally explained by the mildly relativistic motion of the emitting source through a region with a magnetic field gradient, such as a converging flux tube. In our observations, the positive drift rates suggest source motion toward stronger magnetic fields. This differs from the negative drifts often reported in terrestrial auroral kilometric radiation or Jovian S-bursts \citep{Zarka_1996, Zarka_1998, Mutel_2006, Hess_2007}. It also implies that the underlying electron distribution may not require a precipitation-loss process, and could instead be driven by alternative non-loss-cone distributions such as the hollow-beam configuration \citep{Ergun_2000,Treumann_2006}. In such a scenario, the positive drift rate corresponds to a bulk motion of energetic electrons toward stronger magnetic fields, where their initial weakly relativistic parallel energy, possibly augmented by a parallel electric field, is converted into perpendicular energy via magnetic moment conservation in the converging geometry, thereby inducing the positive gradient in the perpendicular velocity distribution essential for ECM growth. We estimated the spatial scale ($\sim$450 km) and traveling distance ($\sim$2.7 Mm) of the radiators by scaling their assumed velocity of $0.3\,c$ with the typical sub-burst duration at a single frequency ($\sim$5 ms) and the total sub-burst duration ($\sim$30 ms), respectively.

Plasma emission, as the other coherent radio mechanism, could also be considered. In this case, the emission frequency is the local plasma frequency or its second harmonic: $f_{\text{pe}} = 8.98\,s\sqrt{n}\,\text{kHz}$, where $n$ is the plasma number density in cm\textsuperscript{-3}. The observed frequencies would correspond to $n=\fromto{1.2}{2.8}\times10^{10}\,\text{cm}^{-3}$ for fundamental emission or $n=\fromto{0.3}{0.7}\times10^{10}\,\text{cm}^{-3}$ for the second harmonic. However, plasma emission is less likely to produce the narrowband, millisecond-scale sub-bursts observed in AD Leo than ECM emission \citep{Vedantham_2021}. Therefore, while plasma emission cannot be entirely ruled out, we consider ECM as the primary emission mechanism in the following discussions.

\subsection{MHD Wave Model for the Modulation of Sub-burst Trains}

\label{sec:model_main}
MHD waves represent one of the most prevalent physical explanations for quasi-periodic phenomena observed in solar and stellar coronae \citep{VanDoorsselaere_2016, McLaughlin_2018,Kupriyanova_2020,Zimovets_2021}. Under linear approximations, these waves can induce sinusoidal temporal variations in plasma density, magnetic field strength, and coronal loop radius, thereby periodically modulating observable parameters such as radiation intensity and frequency \citep{Aschwanden_1987,  Aschwanden_2011, Nakariakov_2020}. Coronal MHD waves are conceptualized as propagating predominantly along the axis of magnetic flux tubes or magnetic slabs \citep[e.g.,][]{Edwin_1983}. This configuration is compatible with the scenario of radiation generated by electrons streaming along magnetic flux tubes. Thus, we propose that MHD waves in AD Leo's corona are a candidate to explain the observed sub-burst train modulations. 

We note here that some of the observed features will fall outside the scope of our simple MHD wave model developed in the following subsections. These include the flux density modulations, the presence of multiple modulated sub-burst chains, the frequency spacing between chains, the growth in frequency modulation amplitude, the overall upward frequency drift in some chains, and the abrupt disappearance of the major chain after about two cycles. These features highlight the richness of the underlying processes and point to directions for future modeling.

\subsection{Examination of MHD Wave Modes}
\label{sec:wave_mode}
This subsection primarily performs quantitative calculations to evaluate the validity of multiple MHD wave modes, based on modulation parameter measurements of the major sub-burst train shown in \figref{fig:mod}. We first present the analysis assuming ECM fundamental emission, supplementing with the results for the second-harmonic case, and discuss the implications. At last, we briefly consider the plasma emission scenario.

The phase velocity of MHD waves relates to the Alfvén speed, the sound speed, or both, depending on the wave mode. Using subscripts i and e to denote internal and external quantities of the flux tube, respectively, and incorporating the low plasma beta condition, a magnetic field strength of $B=B_\text{e}=B_\text{i}=473\,\text{G}$ is derived by using the central frequency of 1325 MHz in \eqnref{eqn:freq_obs}. Adopting the constraint on plasma number density $n_\text{e}<n_\text{i}<10^{10.4}\, \text{cm}^{-3}$, the temperature estimate $2<T<10\, \text{MK}$ \citep{Sanz_2002, Maggio_2004,Duvvuri_2025}, and supplementing with the assumption $n_\text{e} > 10^{9.6}\,\text{cm}^{-3}$ (derived by using a simplified hydrostatic coronal model), we estimate the internal and external Alfvén speeds and the sound speed: 
\begin{align}
    \CAi&<\CAe=B/\sqrt{\mu_0\rho_\text{e}}=B/\sqrt{\mu_0 m_\text{p}n_\text{e}}<16.4\, \text{Mm/s},\\
    \CAe&>\CAi=B/\sqrt{\mu_0\rho_\text{i}}=B/\sqrt{\mu_0 m_\text{p}n_\text{i}}>6.5\, \text{Mm/s},\\
    \Cs&=\sqrt{k_\text{B}T/m_\text{p}}=\fromto{234}{524}\, \text{km/s},
\end{align}
where $m_\text{p}$ is the mass of a proton. 

We quantitatively examine three primary MHD wave modes: fast sausage, fast kink, and torsional Alfvén, considering fluting modes with azimuthal wavenumbers $\left\lvert m \right\rvert > 1$ less probable. Slow modes, while capable of modulating plasma density, are inefficient at modulating the magnetic field under low plasma beta conditions and therefore unlikely to be the direct cause of the observed frequency modulation; they are thus excluded from consideration \citep{Wang_2021}.

For the fast sausage mode, the period $P$ relates to the flux tube radius $a$ via $P = 2\pi a/ (j_{0,l}\, C_{\text{fi}})$, with $C_{\text{fi}} \approx C_{\text{Ai}}$ and $j_{0,l} \approx (l-0.25)\pi$ for the transverse harmonic number $l$ \citep{Kopylova_2007,Li_2020c,Nakariakov_2020}. Using the observed $P = 1.53\,\text{s}$ and $l=1$, we obtain $a = \fromto{3.7}{9.4}\,\text{Mm}$. For non-leaky sausage, fast kink, and torsional Alfvén modes, the phase velocity obeys $C_{\text{Ai}} < C_{\text{p}} < C_{\text{Ai}}$ \citep{Kopylova_2007,Nakariakov_2020,Li_2020c}; for kink modes, $C_{\text{p}}$ is also bounded by the kink speed $C_{\text{K}}$ \citep{Nakariakov_2021, Gao_2025}. These constraints yield wavelengths $\lambda = P C_{\text{p}} = \fromto{9.9}{25.1}\,\text{Mm}$. Assuming a standing-wave resonance in a coronal loop ($L = n\lambda/2$), the fundamental mode ($n=1$) gives loop lengths $L = \lambda/2 = \fromto{5.0}{12.5}\,\text{Mm}$. For the second-harmonic emission case, by using $n_\text{e} > 10^{9.1}\,\text{cm}^{-3}$, similar calculations give $a=\fromto{1.9}{8.3}\,\text{Mm}$ for the leaky sausage mode, and $\lambda=P\Cp=\fromto{5.0}{22.2}\,\text{Mm}$ for the other three modes.

Our derived spatial scales for the various MHD wave modes point to different implications. For instance, the leaky sausage mode is particularly advantageous for generating short-period QPPs within large coronal structures because it requires only a small flux tube radius rather than a short total loop length. It is commonly invoked to explain QPPs on timescales of seconds in the solar corona \citep[e.g.,][]{Carley_2019}. For the other wave modes (non-leaky sausage, kink, and torsional Alfvén), we consider existing estimates of AD Leo's loop sizes. Specifically, spectroscopic studies of AD Leo flares indicate a wide range of possible coronal loop lengths, from compact structures with lengths of a few tenths of the stellar radius \citep[$R_{\text{AD Leo}} \approx 300$ Mm;][]{Kesseli_2019}, to loops comparable to the stellar radius \citep{Hawley_1995, Cully_1997, Reale_1998, Favata_2000, Sciortino_2000, Mullan_2006}. Compared with these estimates, the derived loop length ($L = \lambda/2 = \fromto{5.0}{12.5}\,\text{Mm}$ for $s=1$ and $L=\fromto{2.5}{11.1}\,\text{Mm}$ for $s=2$) for fundamental standing waves is considerably small, implying either very short loops or the presence of higher parallel harmonics ($n>1$) to fit within larger structures. This does not necessarily contradict evidence for larger loops, as a stellar corona likely contains a distribution of loop sizes. Therefore, given the lack of constraints on the specific loop length or harmonic number, we retain all considered wave modes as equally plausible candidates.

Alternatively, if plasma emission is the operative mechanism, the observed frequency modulation would imply density modulation \citep{Yu_2016}. In that case, fast sausage modes and slow modes become relevant, as both can produce periodic density variations. Following a similar estimation presented above, we find that the wavelengths of the two modes are on the order of 0.1 Mm (assuming typical coronal parameters). At present, there is similarly no evidence to exclude such small wavelengths.

\subsection{Simplified Parametric Models for Frequency and Drift Rate Modulations}
\label{sec:param_model}

In general, MHD waves can influence the emission through several pathways: they may modulate the magnetic field strength or its direction, and they can also affect the electron density and velocity distribution via transport and wave-particle interactions. In the ECM framework, the observed frequency modulation is most directly tied to changes in the magnetic field conditions. While both magnetic field strength and direction could contribute to the modulation of the visible radiation frequency, incorporating both effects simultaneously would introduce too many free parameters. Thus, we analyze these two cases separately by developing two distinct parameterized models. 

Within the plasma emission framework, frequency modulation would be attributed to density variations rather than magnetic field changes. The mathematical description of density modulation by an MHD wave is analogous to that of magnetic strength modulation in the ECM case, and a similar parametric forms can be applied. Thus, we do not repeat the analysis in this subsection. Meanwhile, for the sake of simplicity, we will assume the fundamental ECM emission for the subsequent derivations and discussions in this subsection, as the central arguments and conclusions of the models are qualitatively unaffected by a scaling factor. 

\subsubsection{Frequency Modulation by Magnetic Field Strength}
\label{sec:str_mod}
\begin{figure}[]
    \centering
    \includegraphics[width=\textwidth]{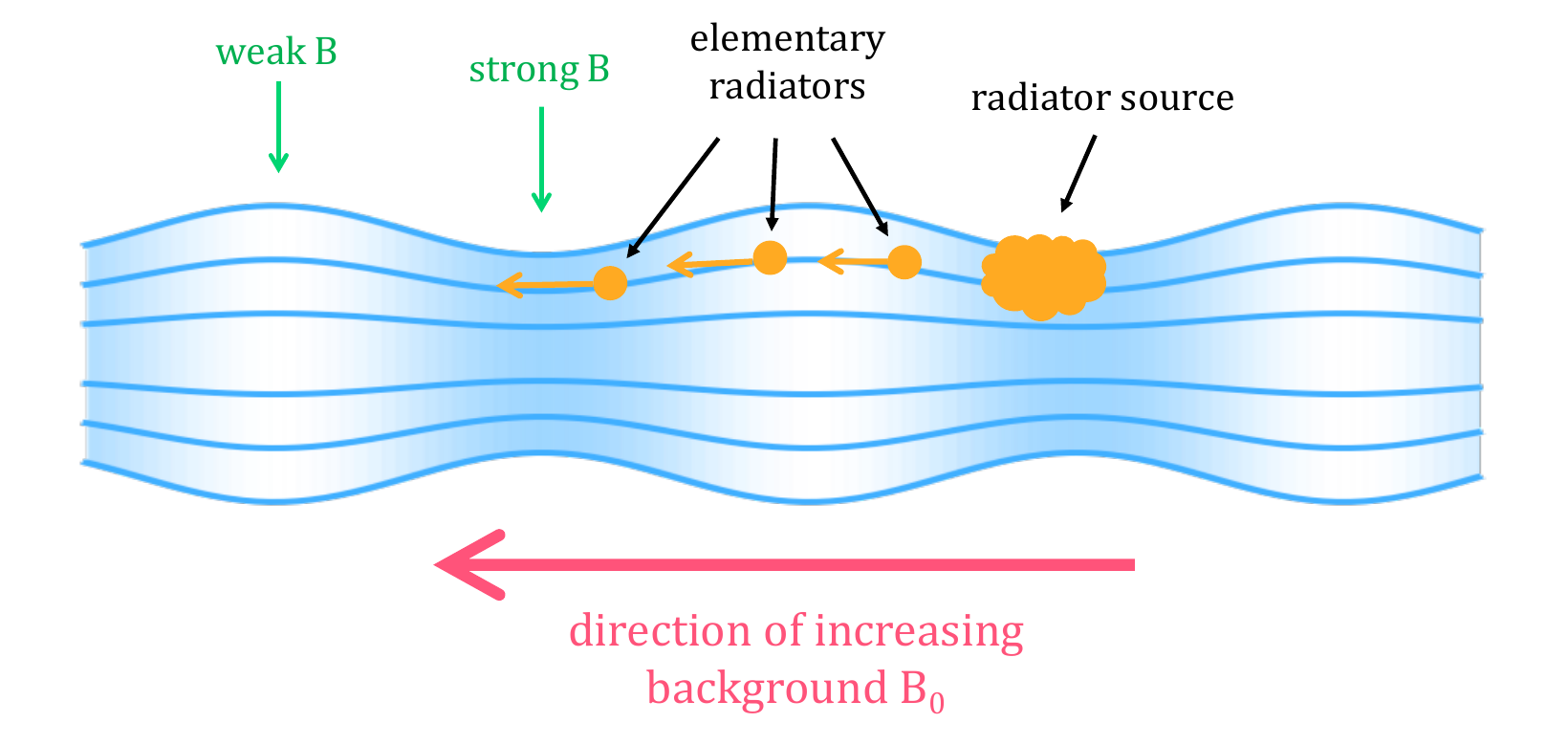}
    \caption{A schematic diagram of an MHD wave modulating the frequency of a sub-burst train through magnetic field strength oscillations. A fast sausage mode wave propagates along in the blue magnetic flux tube, causing the magnetic field to oscillate periodically in both time and space. The red arrow indicates the direction of the background magnetic field gradient along the flux tube axis. The orange cloud pattern represents the injection source of the radiators, while the orange circles represent the ECM elementary radiators moving rapidly along the flux tube. Note that while three radiators are depicted to visualize the trajectory, typically only one radiator is active or visible at any given moment.}
    \label{fig:str_mod}
\end{figure}

Since the frequency of ECM radiation is directly proportional to the magnetic field strength, the change in magnetic field strength caused by a fast sausage mode MHD wave can directly account for the observed frequency variation. As shown in \figref{fig:str_mod}, elementary ECM radiators move within a magnetic flux tube containing a sausage mode MHD wave, and their radiation frequency is modulated by the spatial and temporal changes in the magnetic field strength. For a standing wave and a traveling wave propagating along the flux tube, adopting the exponentially growing sinusoidal form, we can model the magnetic field strength distribution $B(x,t)$ along the flux tube with the following analytical expressions:
\begin{align}
    \begin{split}
        \text{Standing Wave}\ &B(x,t)=B_0(x)+B_1\myexp{t/\tau}\cos(kx+\phi_1)\cos(\omega t+\phi_2);\label{eqn:saus_freq}\\
        \text{Traveling Wave}\ &B(x,t)=B_0(x)+B_1\myexp{t/\tau}\cos(\omega t-kx+\phi_2).
    \end{split}
\end{align}
Here, $x$ is the spatial coordinate along the flux tube, which is also the direction of motion of the ECM radiation source. We assume that $x=0$ is the central point of each radiator's trajectory. $B_0(x)$ represents the background magnetic field strength distribution. By substituting into $f=2.80\,B\, \text{MHz}$ and comparing it at $x=0$ with \eqnref{eqn:freq_obs}, we can easily deduce that $B_0(0)=473\, \text{G}$, $\tau=2.35\, \text{s}$ and $\omega=4.11\, \text{rad/s}$ and $\phi_2=-0.85\pi$. For the standing wave case, $B_1\cos\phi_1=10.2\, \text{G}$, and for the traveling wave case, $B_1=10.2\, \text{G}$. 

The frequency drift rate of a sub-burst is primarily determined by the product of the radiation source's velocity and the background magnetic field gradient. Considering the sub-burst frequency drift rate
\begin{align}
    \diff{f}{t}=2.80\, \diff{B}{t}=2.80\, \left(v\pdiff{B}{x}+\pdiff{B}{t}\right)\, \text{MHz/s}, \label{eqn:drift_theory}
\end{align}
this model predicts:
\begin{align}
    \begin{split}
        \text{Standing Wave}\ \diff{B}{t}=v\pdiff{B_0}{x}&-\left(vk\sin(kx+\phi_1)-\tau^{-1}\cos(kx+\phi_1)\right)B_1\myexp{t/\tau}\cos(\omega t+\phi_2)\\
        &-\omega B_1 \myexp{t/\tau}\cos(kx+\phi_1)\sin(\omega t+\phi_2);\label{eqn:saus_drift}\\
        \text{Traveling Wave}\ \diff{B}{t}=v\pdiff{B_0}{x}&+\tau^{-1}B_1\myexp{t/\tau}\cos(\omega t-kx+\phi_2)\\
        &-(\omega-kv)B_1\myexp{t/\tau}\sin(\omega t-kx+\phi_2).
    \end{split}
\end{align}
The radiator's velocity along the flux tube, $v$, is assumed to be constant here as a simplification. As is evident in the above equations, the drift rate modulations follow the same period and e-fold growth time as the frequency modulations. Thus, the constrained fit of the drift rate modulation in \eqnref{eqn:drift_obs} should be valid in this model.

For the standing wave case, we consider the predicted amplitude of the frequency drift rate modulation (denoted as $D$) in \eqnref{eqn:saus_drift} at $x=0$. 
Substituting the parameters yields 
\begin{align}
    D &=2.80\sqrt{\left(vk\sin\phi_1-\tau^{-1}\cos\phi_1\right)^2+\omega^2\cos^2\phi_1}B_1 \myexp{t/\tau}\,\text{MHz/s}\nonumber\\
    &\geq 2.80\omega B_1\myexp{t/\tau}\cos\phi_1\, \text{MHz/s}\nonumber\\
    &= 117\, \myexp{t/2.35}\, \text{MHz/s}.
\end{align}
This result contradicts the observed value of $D_\text{obs}= 46.0\, \myexp{t/2.35}\, \text{MHz/s}$ in \eqnref{eqn:drift_obs}.  

For the traveling wave case in \eqnref{eqn:saus_freq} and \eqnref{eqn:saus_drift}, the predicted phase difference between the frequency and drift rate modulations is $\delta \phi=\arctan\tau(\omega-kv)$. The measured nearly in-phase ($\delta \phi \approx 0$) behavior is not aligned with the predicted drift rate modulation amplitude $D=2.80\sqrt{\tau^{-2}+(\omega-kv)^2}B_1\myexp{t/\tau}\, \text{MHz}$, which by substituting the values gives:
\begin{align}
    \omega-kv=\pm 1.57\,\text{s}^{-1}.
\end{align}
This would imply a significant phase difference of $\delta \phi=\pm 0.42\pi$.

In summary, we find contradictions between the observed modulations and the predictions of both the standing and traveling wave models. This failure suggests that the phenomenon is not caused by a simple magnetic field strength oscillation from a fast-mode sausage wave assuming ECM emission, or that the models rely on excessive simplifications. However, accounting for a non-constant parallel velocity $v$, which varies due to energy conservation and the first adiabatic invariant, may help reconcile the discrepancy.

\subsubsection{Frequency Modulation by Magnetic Field Direction}
\label{sec:dir_mod}
\begin{figure}[]
\centering
\includegraphics[width=\textwidth]{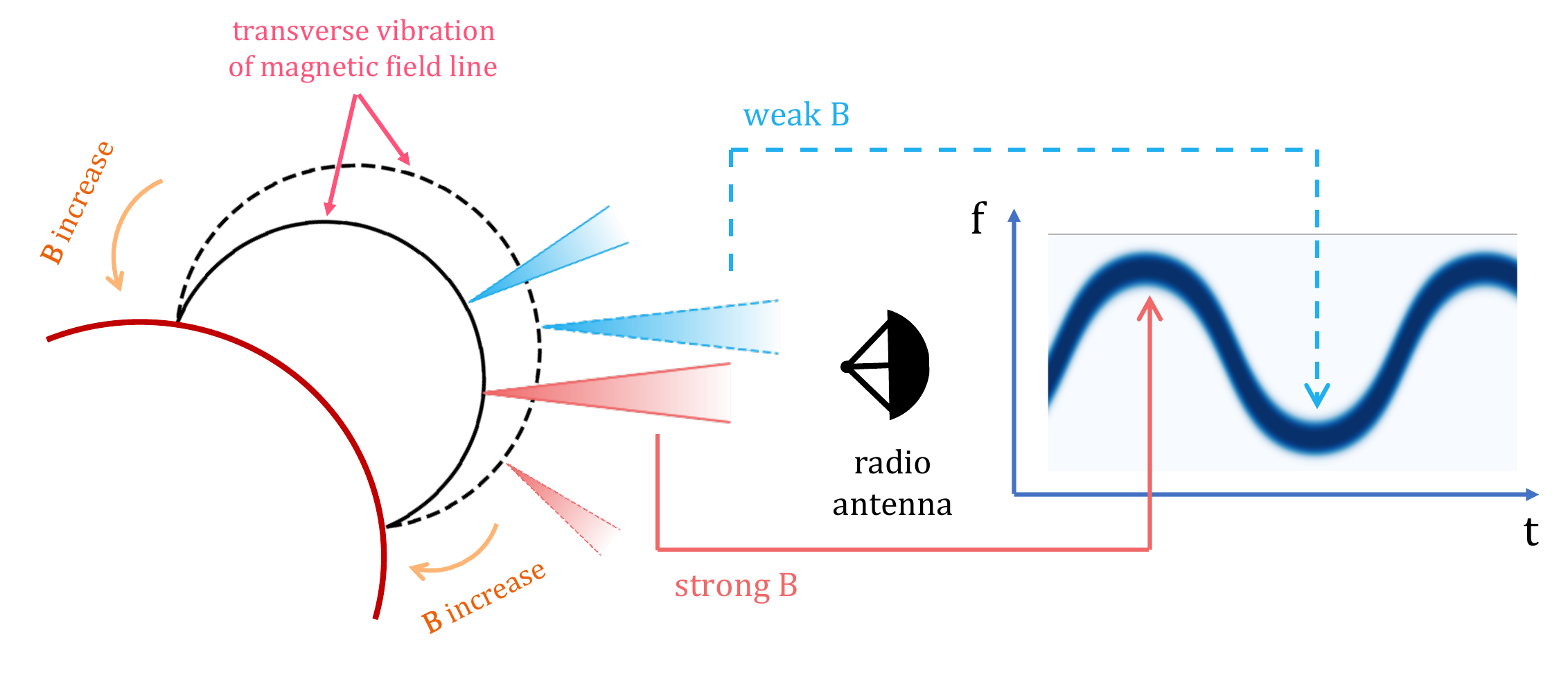}
\caption{A schematic diagram of an MHD wave modulating the frequency of a sub-burst train by changing the magnetic field direction. The figure shows a fundamental mode standing wave in a coronal loop, with black solid and dashed lines representing the laterally oscillating magnetic field lines at different moments. The colored cones represent ECM beams emitted perpendicular to the local magnetic field. Blue cones indicate lower-frequency radiation from the weak-field region near the loop apex; red cones indicate higher-frequency radiation from the strong-field region at lower altitudes. As the loop oscillates, the beams sweep across the line of sight (LOS). Only the beams aligned with the LOS can be observed by the radio telescope, forming the wave-like visibility modulation structure in the dynamic spectrum (right). Note: While this figure depicts in-plane coronal loop vibrations  for clarity, the principle applies equally to out-of-plane vibrations and other magnetic flux tube geometries approximated as an arc.}
\label{fig:dir_mod}
\end{figure}

\begin{figure}[]
\centering
\includegraphics[width=\textwidth]{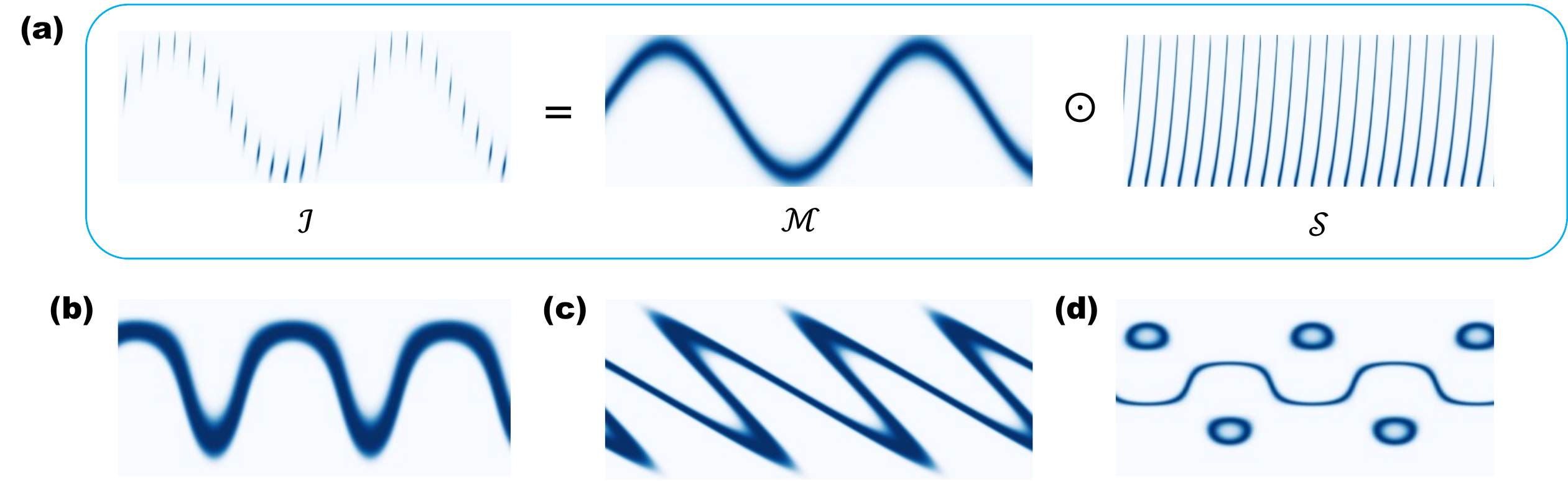}
\caption{Visibility modulation patterns in the simulated dynamic spectra. (a) A visualization of the modulation process, where the final dynamic spectrum is a pixel-wise product (denoted by $\odot$) of the visibility pattern and the linear sub-burst patterns. (b)-(d) Simulated visibility patterns produced by the field direction modulation model described in \secref{sec:dir_mod}. By adjusting the parameters, various dynamic spectrum visibility modulation patterns can be obtained from \eqnref{eqn:visi_cond}, where (b) shows the generation of asymmetric waveforms for a standing wave; (c) shows the hook-like structure that can be produced by a traveling wave with a short wavelength; and (d) shows a case where, with a specific choice of parameters, multiple locations in the flux tube can be visible at the same time for a standing wave. The dynamic spectra in this figure all have time on the horizontal axis and frequency on the vertical axis.}
\label{fig:ds_pattern}
\end{figure}

ECM radiation can be highly directional, emitted within a narrow wall of the emission cone that is quasi-perpendicular to the local magnetic field \citep{Hess_2008a,Hess_2008,Callingham_2024}. Consequently, the visibility of the radio emission depends on the viewing geometry. We propose a model where an MHD wave periodically perturbs the direction of the magnetic field $B$ \citep[e.g.,][]{Tian_2016, Yang_2024, Morton_2025}. While sausage, kink, and torsional Alfvén modes can all involve transverse motions, the collective displacement of a flux tube is most characteristic of a kink-mode oscillation \citep{Nakariakov_2021}, which our model best represents. The perturbation modulates which segment of the flux tube is visible for a given line-of-sight (LOS). As the visible emission point oscillates along the magnetic flux tube, it samples different local magnetic field strengths because of the background field gradient. This change in field strength causes the observed frequency to shift and produces a wave-like pattern in the dynamic spectrum, as illustrated in \figref{fig:dir_mod} and \figref{fig:ds_pattern}(a). The frequency drift rate modulations can arise from a non-zero second spatial derivative of the background field strength \citep{Zhang_2023}.

To formalize the model, we consider a radiator-hosting circular magnetic field line of radius $R$ in the $x$-$y$ plane of a Cartesian coordinate system. The LOS is in the $x$-$z$ plane, making an angle $\theta$ with the $x$-axis, so its unit vector is $\hat{\mathbf{n}}=[\cos\theta,\ 0,\ \sin\theta]$. We consider an MHD wave causing field line displacement purely in the $z$-direction (perpendicular to the loop plane). A position on the unperturbed field line is parameterized by the angle $\phi$ from the $x$-axis. The field line equation is then:
\begin{align}
    \begin{split}
        \text{Standing Wave}\ x&=R\cos\phi,\ y=R\sin\phi,\ z=A\sin{\omega t}\sin(kR\phi+\phi_0);\\
        \text{Traveling Wave}\ x&=R\cos\phi,\ y=R\sin\phi,\ z=A\sin(\omega t-kR\phi).
    \end{split}
\end{align}
Here, $A$ is the amplitude of the field line displacement, $k$ is the wavenumber, and $\phi_0$ is an arbitrary phase; we focuses on producing the wave-like modulation pattern, without considering the amplitude growth feature, which could be incorporated by simply adding a growth term to the amplitude. The differential line element vector $\text{d}\boldsymbol{l}=[\text{d}x,\ \text{d}y,\ \text{d}z]$ can then be calculated. 

Assuming the ECM radiation is emitted perpendicular to the magnetic field, the coordinate $\phi$ of the visible point on the field line satisfies $\hat{\mathbf{n}}\cdot\left(\text{d}\boldsymbol{l}/\text{d}\phi\right)^\text{T}=0$. This yields the visibility condition:
\begin{align}
    \begin{split}
        \text{Standing Wave}\ \sin{\phi}&=kA\tan{\theta}\sin{\omega t}\cos{(kR\phi+\phi_0)};\label{eqn:visi_cond}\\
        \text{Traveling Wave}\ \sin{\phi}&=kA\tan{\theta}\cos{(\omega t-kR\phi)}.
    \end{split}
\end{align}
This indicates that the visible point $\phi$ oscillates periodically in time. Due to the background magnetic field gradient along the loop, the sampled field strength $B$ (and thus the observed frequency $f \propto B$) also oscillates, producing the wave-like envelope in the dynamic spectrum.

In practice, to synthesize a specific visibility pattern $M(t, \phi)$, we evaluate the angle $\alpha$ between the line-of-sight and the local field direction. The visibility is maximal when $\alpha \approx \pi/2$ (emission perpendicular to the field), and we model it with a narrow Gaussian profile of width $\Delta\alpha$ (the ECM beam width). Since $\phi$ can be approximately linearly related to the magnetic field strength $B$ (and hence to frequency $f$) along the gradient, the pattern $M(t, \phi)$ maps directly to a modulation envelope in the $f$-$t$ plane. As is illustrated in \figref{fig:ds_pattern}(a), the final simulated dynamic spectrum is obtained by multiplying this visibility envelope with a base pattern of linear sub-bursts.

We note that by adopting different parameter configurations, \eqnref{eqn:visi_cond} can produce various visibility patterns beyond a symmetric sinusoid (\figref{fig:ds_pattern}(b)-(d)), such as ``hook''-like structures (morphologically similar to some of the Jovian S-bursts discussed by \citet{arkhypov2014dispersion}).


\section{Conclusion}
\label{sec:sum}

This study presents a high-resolution dynamic spectrum observation of radio bursts on AD Leo with the FAST, revealing several wave-like sub-burst trains. Among these, we identified a prominent train with a period of 1.53\,s and a frequency modulation amplitude that grows with a characteristic time of 2.4\,s. Within this train, the central frequency, frequency drift rate, and flux density of the sub-bursts are all modulated. The modulation of the frequency is approximately in-phase with the frequency drift rate modulation and roughly in anti-phase with the flux modulation. The simultaneous modulation of three observable quantities in this event provides a unique opportunity to investigate the modulation mechanism.

Following the work of \citet{Zhang_2023}, we interpret the individual sub-bursts as being generated by fast-moving radiators in a magnetic flux tube. We then analyzed the modulation mechanism of the sub-burst trains, attributing the modulation phenomena primarily to MHD waves in the stellar corona. We found that multiple MHD modes can all in principle account for the short-periodicity of the modulation. Building upon this, we developed two parametric models aiming to reproduce the frequency modulation. The first simplified model relies on a sausage mode MHD wave directly modulating the magnetic field strength in the ECM scenario, but its predicted drift rate modulation is currently inconsistent with the observations. The second model, where MHD waves alter the magnetic field direction and modulate the observed frequency through the LOS effect, provides a promising framework. While pinpointing exact mechanisms remains challenging, this work establishes MHD waves as a plausible interpretation for the wave-like features in the dynamic spectrum, opening new pathways for stellar coronal seismology.

\begin{acknowledgments}
  This work is supported by the National Natural Science Foundation of China grant 12425301 and the Specialized Research Fund for State Key Laboratory of Solar Activity and Space Weather. H. T. is grateful for the support from the New Cornerstone Science Foundation through the Xplorer Prize. This work has used the data from the Five-hundred-meter Aperture Spherical radio Telescope (FAST). FAST is a Chinese national mega-science facility, operated by the National Astronomical Observatories of Chinese Academy of Sciences (NAOC). 
\end{acknowledgments}

\facilities{FAST}

\software{
            NumPy \citep{numpy},
            Matplotlib \citep{matplotlib_2007},
            SciPy \citep{scipy},
            Jupyter \citep{2007CSE.....9c..21P, kluyver2016jupyter},
            Astropy \citep{2013A&A...558A..33A,2018AJ....156..123A,2022ApJ...935..167A}
          }


\bibliography{sample701}{}
\bibliographystyle{aasjournalv7}



\end{document}